\documentclass[10pt]{article}

\usepackage{amsmath}
\usepackage{amssymb}

\usepackage{graphicx}

\usepackage{cite}

\usepackage{color} 

\usepackage[labelfont=bf,labelsep=period,justification=raggedright]{caption}

\makeatletter
\renewcommand{\@biblabel}[1]{\quad#1.}
\makeatother

\date{}

\begin{document}

\begin{flushleft}
{\Large
\textbf{Evolutionary Design in Biological Quantum Computing}
}

Gabor Vattay$^{1,\ast}$, 
Stuart Kauffman$^{2}$
\\
\bf{1} Eotvos University Budapest, Department of Physics of Complex Systems
H-1117 Budapest, Pazmany P. s. 1/A, Hungary
\\
\bf{2} University of Vermont, Vermont Complex Systems Center, 210 Colchester Ave, Farrell Hall, Burlington, VT 05405, USA\\
$\ast$ E-mail: vattay@elte.hu
\end{flushleft}

\section*{Abstract}

The unique capability of quantum mechanics to evolve alternative possibilities in parallel is
appealing and over the years a number of quantum algorithms have been developed offering great  computational benefits. 
Systems coupled to the environment lose quantum coherence quickly and realization of schemes based on
unitarity might be impossible.  Recent discovery of room temperature quantum coherence in light harvesting complexes\cite{Nature.10.1038,Nature08811,Panitchayangkoon20072010,PNAS2011} opens up new possibilities to borrow concepts from biology to use 
quantum effects for computational purposes. While it has been conjectured that light harvesting complexes such as the Fenna-Matthews-Olson (FMO)
complex in the green sulfur bacteria performs an efficient quantum search similar to the quantum Grover's algorithm\cite{Nature.10.1038,Mohseni,Grover} the analogy has yet
to be established.

In this work we show that quantum dissipation plays an essential role in the quantum search performed in the FMO complex and it is fundamentally different from
known algorithms. In the FMO complex not just the optimal level of
phase breaking is present to avoid both quantum localization and Zeno trapping\cite{LloydGuzik,Lloyd} but it can harness quantum dissipation 
as well to speed the process even further up. With detailed quantum calculations taking into account both phase breaking and quantum dissipation we show that
the design of the FMO complex has been evolutionarily optimized and works faster than pure quantum or classical-stochastic algorithms.
Inspired by the findings we introduce a new computational concept based on decoherent  quantum evolution.    
While it is inspired by light harvesting systems, the new computational devices can also be realized on different material basis
opening new magnitude scales for miniaturization and speed.

\section*{Introduction}

In the last five years it became apparent that some biological systems can benefit from quantum effects even at room temperature.
It has been shown experimentally that quantum coherence can stay alive for an anomalously long time in light harvesting complexes\cite{Nature.10.1038,Nature08811,Panitchayangkoon20072010,PNAS2011}. In  these systems excitons initiated by the incoming photons should travel really fast throughout a chain of chromophores in order
to reach the reaction center where they are converted to chemical energy. Excitons decay within 1 nanosecond and dissipate 
energy back to the environment if they cannot find the photosynthetic reaction center via random hopping within that
characteristic time. With classical diffusion via thermal hopping that time is easily consumed, thus evolution should have found more optimal ways to
reach that goal. Quantum mechanics is very helpful in this respect as it allows the system to explore many alternative paths in parallel 
and can discover the optimal one faster than a classical random search would do. However, quantum mechanics has adverse effects too. 
Anderson localization can prevent excitons to travel large distances. Coupling the system to the environment 
breaks phase coherence and can destroy this negative effect of quantum localization. Too much phase breaking however slows down the 
propagation again via the Zeno effect.  At the right amount of phase breaking environmental 
decoherence and quantum evolution collaborate to achieve optimal performance and efficiency. The Environment Assisted Quantum
Transport (ENAQT) theory\cite{LloydGuzik,Mohseni} accounts for the interplay of these two effects and can explain the existence of a transport efficiency optimum at room
temperature relative to both pure quantum or pure classical transport.  ENAQT explains the quick quantum exploration of the search space 
at optimal phase breaking. Once the exciton can reach nearly ergodically the chromophore sites random trapping
delivers of the exciton to the reaction center.

\section*{Results and Discussion}

While ENAQT assures the fast spreading of probability over the light harvesting complex, it  does not guide the exciton to the reaction
center.  The reason for this is that quantum mechanics and phase breaking leads to a uniform probability distribution over the state space.
The reduced density matrix of a system with Hamiltonian $H$ is described by the Lindbad equation\cite{springerlink:10.1007/BF01608499}
\begin{equation}
\partial_t\varrho+\frac{i}{\hbar}\left[H,\varrho\right]=\frac{1}{2}\sum_j \left[V_j\varrho,V_j^+\right]+\left[V_j,\varrho V_j^+\right],
\end{equation}
where the operators $V_j$ describe the coupling of the system and the environment.  In light harvesting systems the Hamiltonian $H_{nm}$ is a discrete,
where the chromophore sites are indexed by $n=1,...,N$. In case the chromophores are
coupled to the environment independently the generators are simply diagonal $V_j=\sqrt{\gamma_{\phi}} \cdot  |j\rangle\langle j|$,
where $\gamma_{\phi}$ is the rate of phase breaking. The Lindblad equation keeps the density matrix normalized during the 
evolution $\mbox{Tr}\{\varrho\}=1$ and its diagonal elements $\varrho_{nn}$ stay positive and give the probability of finding the exciton on
site $n$.  At the optimal level of phase breaking the system relaxes quickly to the uniform probability distribution $\varrho_{nn}=1/N$.
Trapping to the reaction center is described by the imaginary Hamiltonian $-i\hbar \kappa |r\rangle\langle r|$, where $r$ is the site of the reaction center and
$\kappa$ is the trapping rate.  Assuming rapid relaxation to the uniform distribution the bulk of the time an exciton needs to get trapped by the reaction center is determined
by the fraction of time it spends on the chromophore of the reaction center. The reaction center is able to catch an exciton siting on it
in average time $1/\kappa$ and the exciton spends $\varrho_{rr}$ fraction of its time on the chromophore. The average transport time is then the product  
$\langle \tau \rangle \approx 1/(\varrho_{rr}\kappa)=N/\kappa$.  One of the best studied light harvesting systems is the FMO complex\cite{Olson} which consists of $N=7$ chromophores. We use this example thrughout this paper. It has been shown\cite{LloydGuzik} that ENAQT is optimal in this system at phase breaking rates of $\gamma_{\phi}=300 \mbox{cm}^{-1}$ corresponding to room temperature. At trapping rate $1 \mbox{ps}^{-1}$ the exciton needs about 7 ps to reach the reaction center, which is consistent with this
estimate.

Since at optimal phase breaking the transport time depends only on the number of sites and on the trapping rate the concrete form of
the FMO Hamiltonian plays no role as long as the relaxation to the uniform distribution is sufficiently fast. Accordingly, Hamiltonians with
extended wave functions should be slightly more efficient than localized systems since the exciton is not trapped and the relaxation to the uniform distribution is
somewhat faster. 
We demonstrate this in case of the FMO complex where the Hamiltonian $H_{nm}$ has been obtained from 
spectroscopy\cite{Tobias}. The diagonal part of the Hamiltonian consists of the site energies of the chromophores. The off diagonal hopping terms
describe the transition between sites. We can modify the localization properties of this Hamiltonian by rescaling the diagonal elements relative to the
off diagonal elements  $H_{nm}'=H_{nm}+(\lambda-1)\delta_{nm}H_{nn}$ , where $\lambda$ is the tuning parameter. For $\lambda=1$ we recover
the original Hamiltonian. For $\lambda>1$ the diagonal elements become larger and the system becomes completely localized for $\lambda\rightarrow\infty$,
while for $0\leq\lambda<1$ the system becomes more extended. Fig.1 shows the average localization length of the FMO Hamiltonian. It changes
almost monotonically with $\lambda$. 
In Fig. 2 we show the transport efficiency and transport time as a function of $\lambda$ at optimal phase breaking calculated with the parameters of Ref. 5.
The details are in the supplementary material. 
Both of them change monotonically with $\lambda$ and the transport is slightly more efficient and faster for the delocalized case as we expected.  
The real FMO complex at $\lambda=1$ is not optimal in any sense. As we show next, evolution optimized the transport process further by 
guiding the excitons to the reaction center which sits at the lowest energy site and the design of the FMO complex is in fact highly optimal.
To show this we have to go beyond the Lindblad equation in order to account for the relaxation to thermal equilibrium.

One way to study the relaxation to the correct thermal equilibrium is to use the Redfield equations describing the interaction of the system and the environmental bath. 
The Redfield equation can be cast in a form similar to the Lindblad equation\cite{Thomas,Kohen}
\begin{equation}
\partial_t\varrho+\frac{i}{\hbar}\left[H,\varrho\right]=\sum_j \left[V_j^+\varrho,V_j\right]+\left[V_j,\varrho V_j^-\right],\label{redfield}
\end{equation} 
where the operators  can be written in energy representation as  $\left[V^+_j\right]_{ab}=\left[V_j\right]_{ab}/(1+e^{\beta(E_a-E_b)})$ and $\left[V^-_j\right]_{ab}=\left[V_j\right]_{ab}/(1+e^{-\beta(E_a-E_b)})$. The operators coupling the bath and the environment are physical observables hence self-adjoint $V_j=V_j^+$.
The equilibrium solution of this equation is the Boltzmann distribution $\varrho=\exp(-\beta H)/Z$, where $Z=\mbox{Tr}\{\exp(-\beta H)\}$ is the partition function. The uniform distribution
is recovered for infinite temperature $\beta=0$. 

For high temperatures we can expand this equation for small $\beta$. The first two terms in the expansion are basis independent
\begin{equation}
\partial_t\varrho+\frac{i}{\hbar}\left[H,\varrho\right]=\frac{1}{2}\sum_j \left[V_j,\left[V_j,\varrho\right]\right]+\frac{\beta}{2}\left[V_j, \{ \left[H,V_j \right], \varrho \} \right],\label{neweq}
\end{equation}
while the third term in the expansion is zero in general as we show in the supplementary material.
The first term is the Lindblad equation for self-adjoint operators $V_j$. The second term describes quantum dissipation, which is missing from the Lindblad equation.
Caldeira and Leggett (CL) showed that the reduced density matrix of open quantum systems coupled to a high temperature bath experience both phase breaking and quantum dissipation  and satisfy the equation
\begin{equation}
\partial_t\varrho+\frac{i}{\hbar}\left[H,\varrho\right]=\frac{1}{2}\gamma_{\phi}\left[x,\left[x,\varrho\right]\right]-\frac{i\hbar\beta}{2m }\gamma_{\phi}\left[{x},\{{p},\varrho\}\right].
\end{equation}
Our new equation (\ref{neweq}) gives back the CL equation as a special case for the Hamiltonian $H(x,p)=\frac{1}{2m}{ p}^2+U({x})$ with coupling $V=\sqrt{\gamma_{\phi}}{ x}$ and it is valid  
for a much larger class of Hamiltonians and operators $V$.  In particular for discrete Hamiltonians $H_{nm}$ describing the exciton dynamics in light harvesting complexes and
for environmental couplings $V_j=\sqrt{\gamma_{\phi}} \cdot  |j\rangle\langle j|$ it takes the form
\begin{eqnarray}
\partial_t\varrho_{nm}+i\left[H,\varrho\right]_{nm}&=&-2\gamma_{\phi}(1-\delta_{nm})\varrho_{nm}-(1-\delta_{nm})\frac{\gamma_{\phi}\beta}{2}\{H,\varrho\}_{nm}\\ \nonumber
&-&\frac{\gamma_{\phi}\beta}{2}\left(H_{nm}\varrho_{mm}+\varrho_{nn}H_{nm}-H_{nn}\varrho_{nm}-\varrho_{nm}H_{mm}\right).\\ \nonumber
\end{eqnarray}
The most important feature of this equation is that the quantum dissipative term cannot be chosen arbitrarily in models of exciton dynamics. The Hamiltonian and the generators $V_j$
determine both phase breaking and dissipation uniquely.  Also the order of magnitude the dissipative term relative to the phase breaking term is determined by the
ratio of the size of the typical Hamilton matrix element and the temperature. In light harvesting systems these are comparable and quantum dissipation cannot be neglected. 

Quantum dissipation speeds up the transport process in light harvesting complexes. If the site energies at the reaction center are lower than in the other 
parts of the complex the equilibrium density is higher and the exciton spends longer time on the chromophore related to the reaction 
center and is trapped with higher probability. The average time is again $\langle \tau \rangle =1/(\kappa\varrho_{rr})$ but now the probability is
given by the Boltzmann factor $\varrho_{rr}= \langle r| e^{-\beta H}|r\rangle/Z$. In case of the FMO complex this probability is about $40\%$ and
 the transport time would drop to a mere $2.5$ picoseconds in this approximation at optimal phase breaking.  Our detailed calculation using the 
Redfield operators outlined in the supplementary material yields about $3.5$ picoseconds which is very close to this estimate and less than half than it would be without quantum dissipation. We can now ask in what sense is this result optimal? Could we  achieve a better result by choosing as deep site energy
as possible so that $\varrho_{rr}\approx 1$ can be achieved? We show next that this absolute optimum cannot be attained and  the real FMO operates with the best transport time 
possible physically and evolutionarily.

At the optimal phase breaking of ENAQT quantum dissipation introduces a tread-off between fast relaxation to the equilibrium distribution and 
the shape of the equilibrium distribution.  The equilibrium density matrix can be expressed in terms of the energy eigenstates $\psi_n^{(k)}$ as
\begin{equation}
\varrho_{nn}=\sum_k |\psi_n^{(k)}|^2\frac{e^{-\beta E_k}}{Z}.
\end{equation}   
If the system is completely delocalized the wave functions are extended $|\psi_n^{(k)}|^2\approx 1/N$ and the diagonal elements of the density matrix become uniform
$\varrho_{nn}\approx 1/N$ independently of the energy levels $E_k$ of the system.  In this case the relaxation to the equilibrium is fast since the extended wave functions overlap
strongly with the exciton starting on one of the chromophores, but the exciton spends time on each chromophore nearly equally.
If the system is strongly localized the wave functions are concentrated on single sites
$|\psi_n^{(k)}|\approx \delta_{nk}$ and $\varrho_{nn}\approx e^{-\beta E_n}/Z$, where the energy levels are close to the site energies  $E_n\approx H_{nn}$. 
To have localization the site energies should be much larger than the hopping terms in the Hamiltonian. In equilibrium the exciton would spend long time in the neighborhood
of the chromophore with the lowest site energy, but the relaxation time to this equilibrium is very large.
The overlap of the wave function localized on the
lowest energy site with the initial site of the exciton is very small and the exciton stays localized near to its entry site for a very long time.
In Fig. 2 we show both the transport efficiency and transport time for the FMO complex at the optimal phase breaking for different $\lambda$-s tuning
the localization length of the system. For $\lambda>1$ we see a fast drop of transport efficiency and increase of transport time due 
to the slow relaxation hampered by the localization of the exciton. For $\lambda<1$ we see also a monotonic drop of efficiency due to the flattening of
the equilibrium distribution. The shortest transport time and highest efficiency is near the real FMO complex $\lambda\approx 1$, where the states
are neither too localized nor too much extended and realize the tread-off. Note, that the Hamiltonian
is reconstructed from experiments and it carries some level of error.  In Fig. 1 we can see that the
localization length of the real FMO is just half way between the fully localized case, where the wave functions are concentrated on a single site and
the maximally delocalized case, where the states are spread  the most. 

We think that this picture is quite general. If we consider larger transport systems the optimum would again lie somewhere midway between 
the extended and localized cases. Since the localization-delocalization transition is getting sharper with increasing system size these systems can
only be found at parameters near the metal-insulator threshold. To demonstrate this in Fig. 3 we show the the transport efficiency for the golden mean Harper 
model which is one of the simplest models on which the metal-insulator transition can be studied\cite{PhysRevLett.84.1643}. In this model we can see qualitatively the same
behavior and an optimal transport near the localization delocalization (or metal-insulator) transition. It is important to note that even in this large
system the transport time at the optimal phase breaking is still determined by the shape of the equilibrium distribution and the relaxation time is negligible.
It seems advantageous for biologically relevant quantum transport to tune the system into the critical point of the localization-delocalization transition.

How could we use this mechanism to build new types of computers? In the light harvesting case the task of the system is to transport
the exciton the fastest possible way to the reaction center whose position is known. In a computational task we usually would like to find the
minimum of some complex function $f_n$. For the simplicity let this function have only discrete values from $0$ to $K$.
If we are able to map the values of this function to the electrostatic site energies of the chromophores $H_{nn}=\epsilon_0 f_n$ and 
we deploy reaction centers near to them trapping the excitons  with some rate $\kappa$ and can access the current at each reaction center
it will be proportional with the probability  to find the exciton on the chromophore $j_n\sim \kappa\varrho_{nn}$. Since the excitons 
will explore the Boltzmann distribution the currents will reflect that $j_n=\kappa \langle n| e^{-\beta H}|n\rangle/Z$. 
There are three conditions which should be valid simultaneously: 1, The system should operate at the optimal phase breaking which then
should be in the order of magnitude of the energy steps
$\gamma_{\phi}\sim {\cal O}( \epsilon_0)$. 2, In the worst case scenario the minimum current is elevated with a factor   $e^{\beta\epsilon_0}$
relative to the second smallest minimum. To be able to detect this the energies should be of the order of the thermal energy 
$\epsilon_0\sim {\cal O} (k_BT)$. 3, The hopping terms $H_{nm}$ between the chromophores should be optimal to keep the system at
the border of the localization-delocalization transition. The first two conditions can be easily met since the phase breaking is usually 
of the same order as the thermal energy $\gamma_{\phi}\sim k_BT$. The third condition can be realized by placing the chromophores
interacting via the dipole interaction to an optimal distance from each other randomly so that the quasy random $H_{nm}$ matrix elements
keep the system at the localization-delocalization threshold. Conversely, given a random arrangement of $H_{nm}$-s the parameter
$\epsilon_0$ can be tuned so that the system gets to the localization-delocalization threshold. 

This quantum-classical optimization method discovered by evolution seems to be superior to the optimization methods developed so far. Classical
stochastic optimization techniques can be trapped in local minima for long times and careful annealing techniques are required to reach the correct
minimum. Even then sites are discovered in a classical sequential manner and it takes the process long times to find the minimum.  
Quantum mechanics is more advantageous as it is able to explore the sites in parallel, but the discovery process is hampered by Anderson localization
especially near local minima. An optimal amount of phase breaking can destroy the interferences causing this and can ensure the ergodic
exploration of the states while quantum dissipation takes all the advantages of the classical stochastic optimization and establishes the Boltzmann distribution
which elevates the proper minimum. The physical speed of the process is determined by the inverse trapping rate $1/\kappa$ which is in the order of picoseconds.

Current computers operate with about 4 GHz processors, where the cycle time of logical operations is 250 picoseconds. Computers based on artificial light
harvesting complexes could have units with 100-1000 times larger efficiency at room temperature. But, it is also possible to realize such systems on excitons
of organic molecules or on Hamiltonians arising in nuclear matter, which would provide a virtually endless source of improvement both in time and miniaturization
below the atomic scale. Since the realization of this mechanism seems now relatively easy, it is an important question if it has been realized in light
harvesting systems or is also present in other biological transport or optimization processes. Especially in the human brain\cite{Kauffman}.

\section*{Materials and Methods}

\subsection{Redfield Equations for Environment Assisted Quantum Transport}

The Redfield equation can be cast into a form similar to the Lindblad equation (see W. T. Pollard and R. A. Friesner, J. Chem. Phys. 100, 5054 (1997)). In energy representation:
\begin{equation}
\partial_t\varrho_{ab}+\frac{i}{\hbar}\left[H,\varrho\right]_{ab}=\sum_j \left[V_j^+\varrho,V_j\right]_{ab}+\left[V_j,\varrho V_j^-\right]_{ab},\label{RFE}
\end{equation} 
where the operators  can be written in energy representation as  $\left[V^+_j\right]_{ab}=\left[V_j\right]_{ab}/(1+e^{\beta(E_a-E_b)})$ and $\left[V^-_j\right]_{ab}=\left[V_j\right]_{ab}/(1+e^{-\beta(E_a-E_b)})$. The operators coupling the bath and the environment are physical observables hence self-adjoint $V_j=V_j^+$.
The equations can be written also in the form of 
\begin{equation}
\partial_t\varrho_{ab}+\frac{i}{\hbar}\left[H,\varrho\right]_{ab}=\sum_{cd}R_{abcd}\varrho_{cd},
\end{equation}
where
\begin{equation}
R_{abcd}=\frac{V_{ac}V_{db}}{1+e^{\beta(E_a-E_c)}}-\sum_i\frac{V_{ai}V_{ic}\delta_{db}}{1+e^{\beta(E_i-E_c)}}+\frac{V_{ac}V_{db}}{1+e^{-\beta(E_d-E_b)}}
-\sum_i \frac{V_{ai}V_{ic}\delta_{bd}}{1+e^{-\beta(E_a-E_i)}}.
\end{equation}
The energy representation of the coupling operator is $V_{ab}^j=\sqrt{\gamma_{\phi}}{\psi^{a}_j}^*{\psi}_j^b$, where $\psi_j^a$ is the energy eigenstate corresponding to $E_a$ and site index $j$. We can then transform back the equations into site representation and can carry out the summations for $j$ yielding
\begin{equation}
\partial_t\varrho_{nm}+\frac{i}{\hbar}\left[H,\varrho\right]_{nm}=\gamma_{\phi}\sum_{kl}K_{nmkl}\varrho_{kl},
\end{equation}
where
\begin{equation}
K_{nmkl}=A^+(n,k,m)\delta_{ml}+\delta_{nk}A^-(m,l,n)-A^+(n,k,n)\delta_{ml}-\delta_{nk}A^-(m,l,m),
\end{equation}
and
\begin{equation}
A^{\pm}(n,m,j)=\sum_{ab}\frac{{\psi^*}_n^{a}\psi^a_j{\psi^*}^b_j\psi^b_m}{1+e^{\pm\beta(E_a-E_b)}}.
\end{equation}

\subsection{The generalized Caldeira-Legget equations}

We can expand the operators in the Redfield equations in energy representation up to the second power of $\beta$ as
\begin{equation}
\frac{\left[V_j\right]_{ab}}{1+e^{\beta(E_a-Eb)}}=\frac{1}{2}\left[V_j\right]_{ab}-\frac{\beta}{4}(E_a\left[V_j\right]_{ab}-\left[V_j\right]_{ab}E_b).
\end{equation}
Note that the $\beta^2$ term is identiaclally zero. The second term is independent of the representation and can be written as
\begin{equation}
V^\pm_j=\frac{1}{2}V_j-\pm\frac{\beta}{4}\left[H,V_j\right],
\end{equation}
where we use the commutator $\left[H,V_j\right],=HV_j-V_jH$. Substituting this into the Redfield equation (\ref{RFE}) yields
\begin{equation}
\partial_t\varrho+\frac{i}{\hbar}\left[H,\varrho\right]=\frac{1}{2}\sum_j \left[V_j,\left[V_j,\varrho\right]\right]+\frac{\beta}{2}\left[V_j, \{ \left[H,V_j \right], \varrho \} \right].
\end{equation}

\subsection{The Harper model}

The Harper model is defined by the one dimensional chain with site energies $H_{nn}=2\lambda J\cos(2\pi G n)$ and hopping terms $H_{n,n+1}=J$, where $G=(\sqrt{5}-1)/2$ is
the golden mean and $\lambda$ is the tuning parameter. If $\lambda=1$ the system is at the critical point of the localization-delocalization transition. For $\lambda>1$ all the states are
localized in an infinite system and for $\lambda<1$ they are all extended. Fixing the external temperature at $277K$ and the phase breaking at $300 cm^{-1}$ (in spectroscopic wavenumber units) similar to the FMO complex
the parameter $J$ defines the energy scale of the model. For values $J\approx 100-1000 cm^{-1}$ the phase breaking seems to be optimal in a chain of length $N=30$ and 
the best efficiency and the smallest transport time is attained at the critical point of the metal-insulator transition at $\lambda=1$.

\subsection{Calculation of the efficiency and transport time}

Transport time and efficiency calculations coincide with those presented in Patrick Rebentrost, Masoud Mohseni, Ivan Kassal,
Seth Lloyd and Al‡n Aspuru-Guzik, {\em Environment-assisted quantum transport}, New Journal of Physics 11 (2009) 033003.
The same trapping rate $\kappa=1ps^{-1}$ and exciton decay rate $\Gamma=1 ns^{-1}$ is used throughout this paper.
The transport efficiency and transport time is calculated with the numerical inversion of the superoperator discussed in 
M. Mohseni, P. Rebentrost, S. Lloyd and A. Aspuru-Guzik, {\em Environment-assisted quantum walks in photosynthetic energy transfer},
J. Chem. Phys. 129 174106 (2008).

\section*{Acknowledgments}

\bibliography{Evdesign}

\begin{thebibliography}{}
\expandafter\ifx\csname url\endcsname\relax
  \def\url#1{\texttt{#1}}\fi
\expandafter\ifx\csname urlprefix\endcsname\relax\def\urlprefix{URL }\fi
\providecommand{\bibinfo}[2]{#2}
\providecommand{\eprint}[2][]{\url{#2}}


\bibitem{Nature.10.1038}
Engel G.S,, Calhoun T.S., Read E.L., Ahn T-K., Mancal T., Cheng Y-C., Blankenship R.E. and Fleming G.R.\\
 Evidence for wavelike energy transfer through quantum coherence in photosynthetic systems \\
 Nature \textbf{446} (2007), 782 -- 786.
  
\bibitem{Nature08811}
Collini E., Wong C-Y., Wilk K.E., Curmi P.M.G, Brumer P. and Scholes G.D.,\\
Coherently wired light-harvesting in photosynthetic marine algae at ambient temperature\\
Nature \textbf{463} , 644 -- 647 (2010).

\bibitem{Panitchayangkoon20072010}
Panitchayangkoon G., Hayes D., Fransted K.A.,  Caram J.R., 
  Harel E., Wen J., Blankenship R. E.  and Engel G.S.\\
Long-lived quantum coherence in photosynthetic complexes at physiological temperature\\ 
Proceedings of the National Academy of Sciences  \textbf{107} no.~29, 12766--12770. (2010)

\bibitem{PNAS2011}
Panitchayangkoon G., Voronine D.V., Abramavicius D., Caram J.R., Lewis N.H.C., Mukamel S., and Engel G.S.\\
Direct evidence of quantum transport in photosynthetic light-harvesting complexes\\
PNAS 2011 ; published ahead of print December 13 (2011)

\bibitem{LloydGuzik}
Patrick Rebentrost, Masoud Mohseni, Ivan Kassal,
Seth Lloyd and Al‡n Aspuru-Guzik\\
{\em Environment-assisted quantum transport}\\
New Journal of Physics 11 (2009) 033003

\bibitem{Mohseni}
M. Mohseni, P. Rebentrost, S. Lloyd and A. Aspuru-Guzik,\\
 {\em Environment-assisted quantum walks in photosynthetic energy transfer}\\
J. Chem. Phys. 129 174106 (2008)

\bibitem{Grover}
Grover L.K\\
A fast quantum mechanical algorithm for database search, \\
Proceedings, 28th Annual ACM Symposium on the Theory of Computing, (May 1996) p. 212

\bibitem{Lloyd}
S. Lloyd\\
Quantum coherence in biological systems\\
Journal of Physics: Conference Series 302, 012037 (2011)

\bibitem{CaldeiraLeggett}
A. O. Caldeira and A. J. Leggett\\
Path integral approach to quantum Brownian motion\\
Physica A {\bf 121}, 587-616 (1983)


\bibitem{Olson}
Olson J.M.\\
The FMO protein\\
Photosynthesis Research 80: 181Ð187, 2004.


\bibitem{springerlink:10.1007/BF01608499}
Lindblad G.\\
On the generators of quantum dynamical semigroups\\
Communications in Mathematical Physics \textbf{48} (1976), 119--130 (1976)

\bibitem{Tobias}
Minhaeng Cho, Harsha M. Vaswani, Tobias Brixner, Jens Stenger, and Graham R. Fleming\\
Exciton Analysis in 2D Electronic Spectroscopy\\
Phys. Chem. B, 2005, 109 (21), pp 10542-10556 (2005)




\bibitem{Thomas}
W. Thomas Pollard and Richard A. Friesner\\
Solution of the Redfield equation for the dissipative quantum dynamics of multilevel systems\\
J. Chem. Phys. 100, 5054 (1994)

\bibitem{Kohen}
D. Kohen, C. C. Marston, and D. J. Tannor\\
Phase space approach to theories of quantum dissipation\\
J. Chem. Phys. 107, 5236 (1997)

\bibitem{PhysRevLett.84.1643}
Evangelou, S. N. and Pichard, J.-L.\\
Critical Quantum Chaos and the One-Dimensional Harper Model\\
Phys. Rev. Lett.  {\bf 84}, 8, 1643--1646 (2000)

\bibitem{Kauffman}
Kauffman S.A.\\
Answering Descartes: Beyond Turing\\
In S. Barry Cooper and Andrew Hodges (editors) "The Once and Future Turing: Computing the World", Cambridge University Press (2012)\\ 
reprinted in Proceedings of the Eleventh European Conference on the Synthesis and Simulation of Living Systems
Edited by Tom Lenaerts, Mario Giacobini, Hugues Bersini, Paul Bourgine, Marco Dorigo and RenŽ Doursat, MIT Press (2011)\\
\verb+\http://mitpress.mit.edu/books/chapters/0262297140chap4.pdf+


\end{thebibliography}

\section*{Figure Legends}

\begin{figure}[!ht]
\begin{center}
\includegraphics[width=6in]{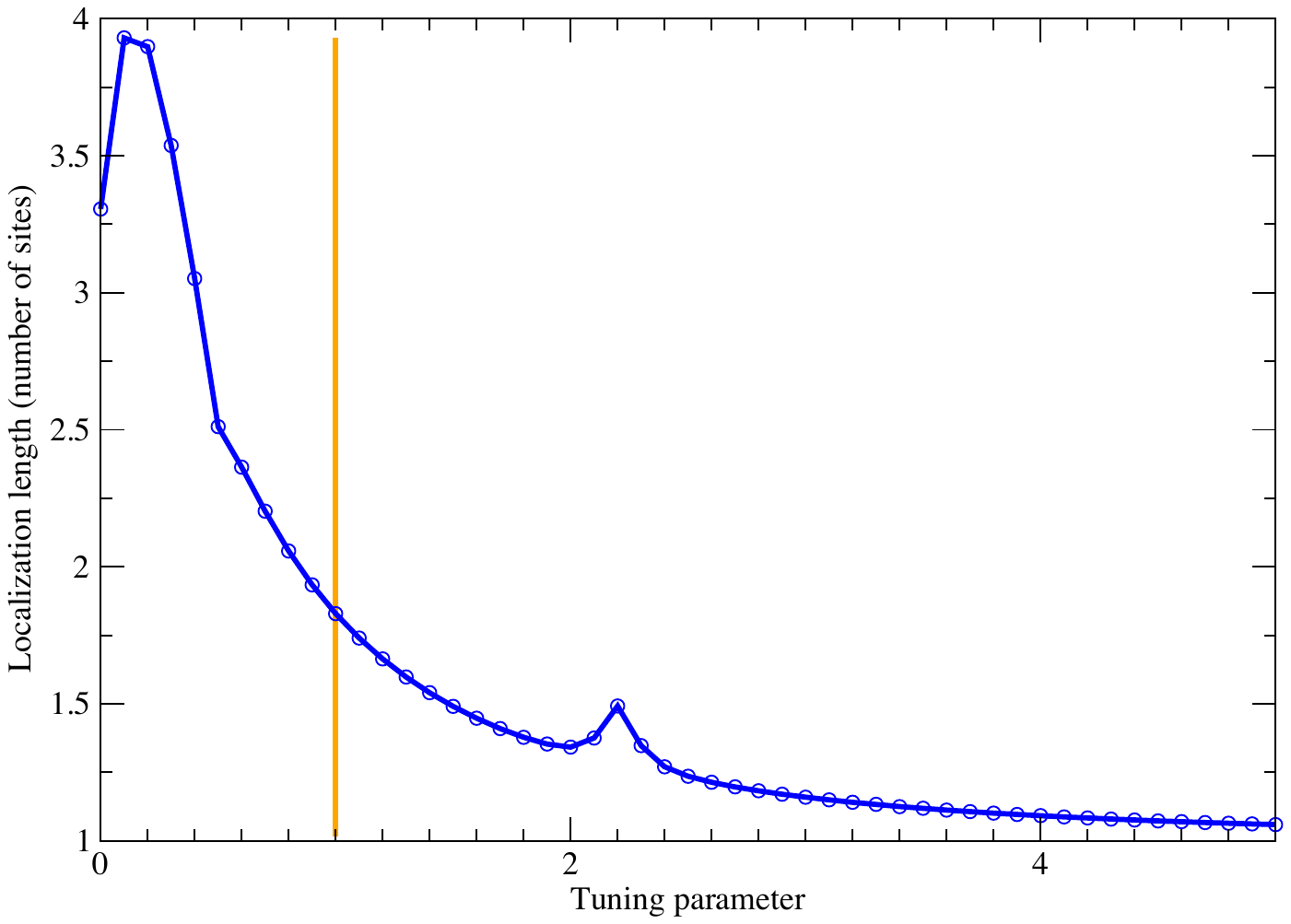}
\end{center}

\caption{
Average localization length of the FMO complex as a function of the tuning parameter $\lambda$. The value $\lambda=1$ corresponds
to the real FMO complex. For $\lambda>1 (<1)$ the diagonal elements of the Hamiltonian matrix are magnified (shrinked) causing more (less)
localization.  The localization length is the reciprocal of the inverse participation ratio $\xi=1/\mbox{IPR}$
calculated as an average for all the $N=7$ eigenfunctions $(k)$ and sites $n$ of the FMO complex $\mbox{IPR}=( \sum_{n,k=1}^N |\psi_n^{(k)}|^4)/N$.
The localization length shows how many sites are involved in a given energy eigenfunction in average. Conversely it also shows how many
energy eigenstates overlap in a given site. The value $\xi=1$ means that the energy eigenfunctions are localized on a single state while
$\xi=4$ seems to be the largest level of attainable delocalization.  The FMO complex is half way between the fully localized and delocalized cases. }
\end{figure}

\begin{figure}[!ht]
\begin{center}
\includegraphics[width=6in]{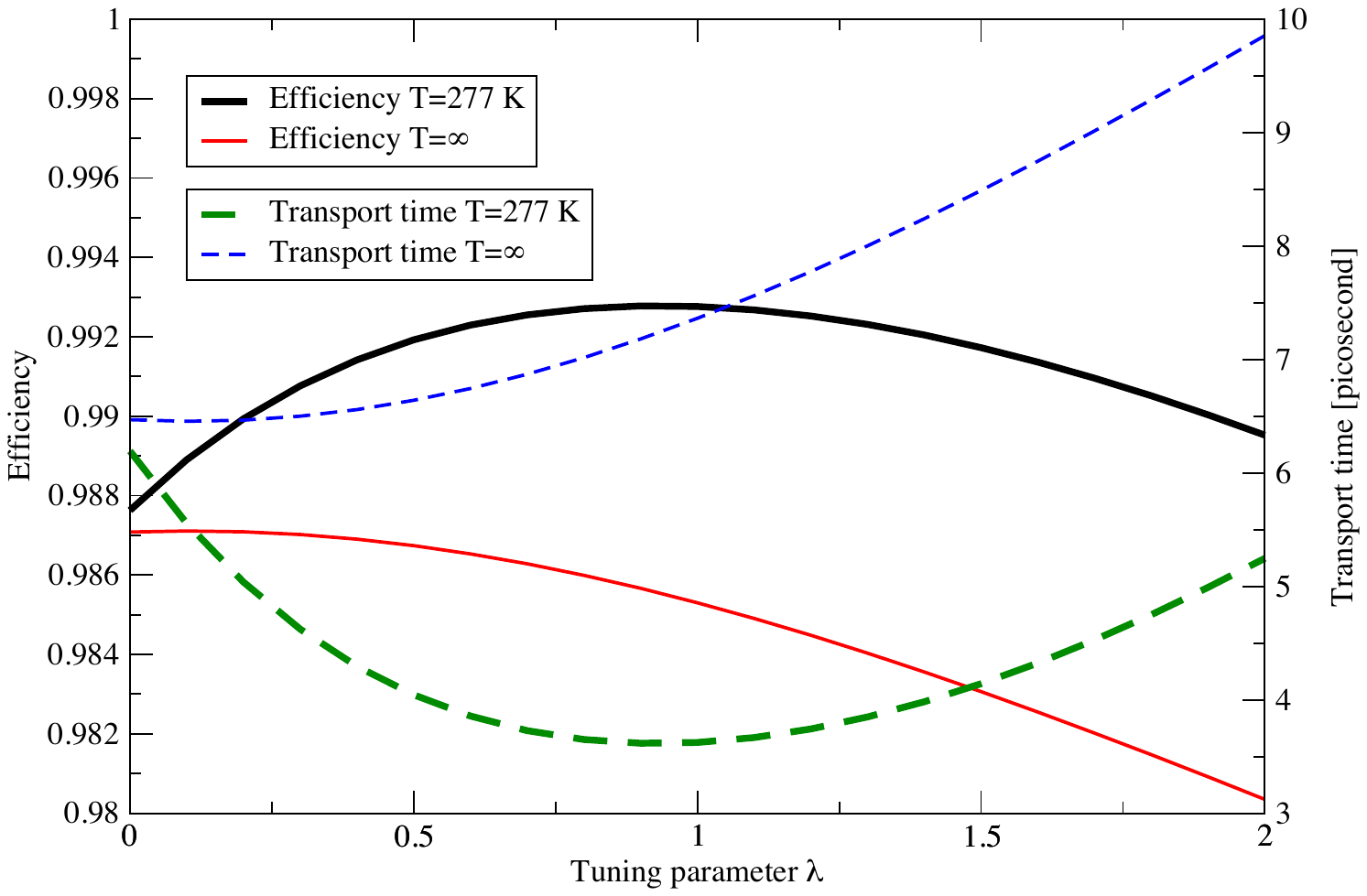}
\end{center}
\caption{Transport efficiency and transport time in the FMO complex as a function of the tuning parameter $\lambda$ at optimal phase breaking $\gamma_{\phi}=300 cm^{-1}$.  
Solid curves show transport efficiency for ambient temperatures $T=277$ (black) and for the case when quantum dissipation is not present $T=\infty$ (red). 
The presence of quantum dissipation increases the transport efficiency for all parameters $\lambda$. Without quantum dissipation the delocalized systems
$\lambda<1$ are more efficient than the localized ones $\lambda>1$ and efficiency increases with the localization length. When quantum dissipation is
present the efficiency increases with about $0.8\%$ in the optimal point near $\lambda\approx 1$ and shows a maximum near the real FMO complex. (Note that
the experimental parameters of the FMO Hamiltonian carry some error and the maximum cannot be expected exactly at $\lambda=1$.)
Dashed lines show the transport time. There is a dramatic speedup of transport due to quantum dissipation. The transport time drops from about 7 picoseconds to
3.5 picoseconds for the FMO complex. The transport time without quantum dissipation (blue) changes monotonically with the localization and fastest for the
most delocalized case. With quantum dissipation (green) the transport time is about minimal for the real FMO complex which is in between the localized and delocalized
cases.}
\end{figure}

\begin{figure}[!ht]
\begin{center}
\includegraphics[width=6in]{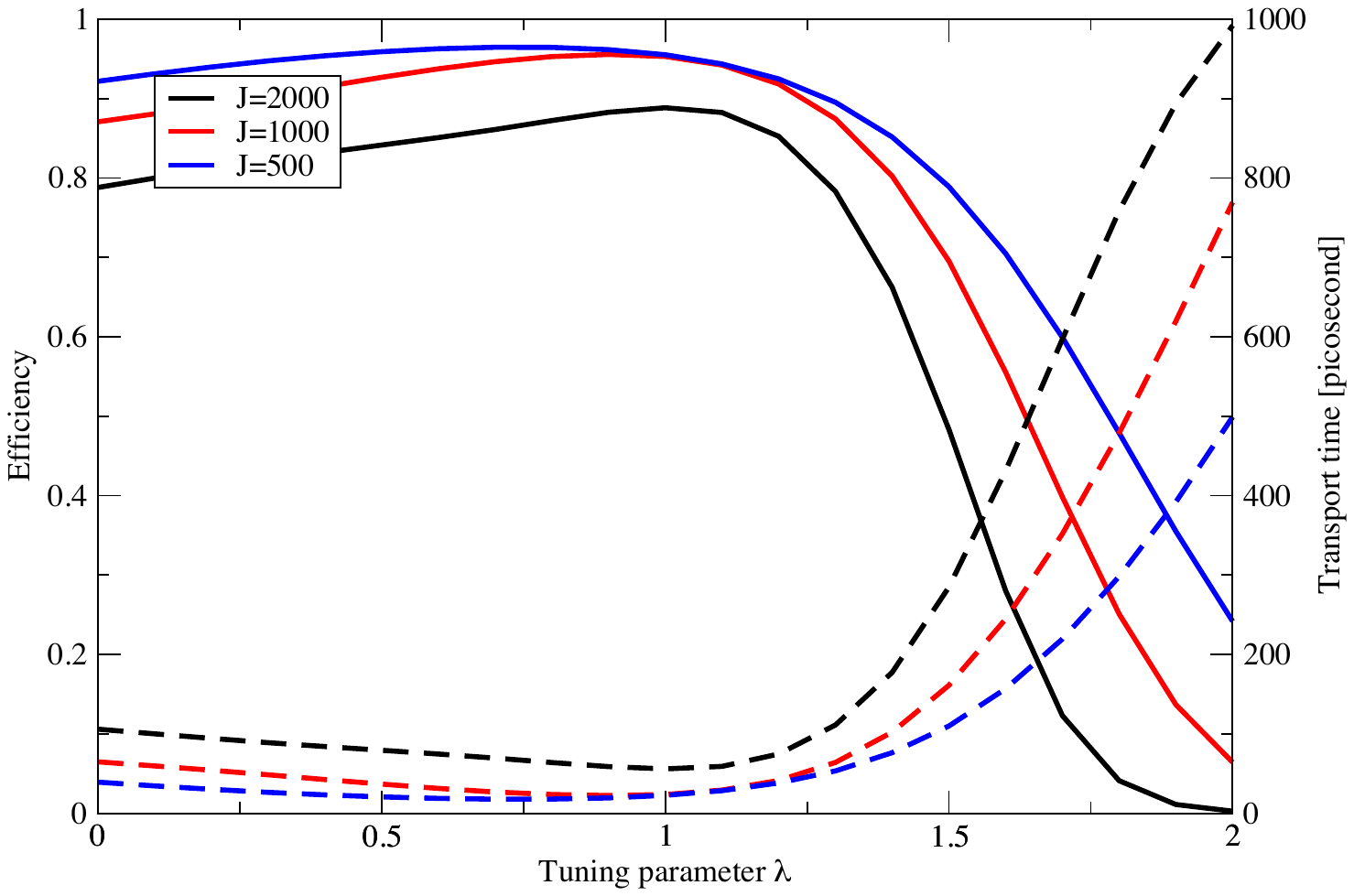}
\end{center}
\caption{Transport efficiency and transport time for the golden mean Harper model as a function of the tuning parameter. The Harper model is defined by the one dimensional chain with site energies $H_{nn}=2\lambda J\cos(2\pi G n)$ and hopping terms $H_{n,n+1}=J$, where $G=(\sqrt{5}-1)/2$ is
the golden mean and $\lambda$ is the tuning parameter. If $\lambda=1$ the system is at the critical point of the localization-delocalization transition\cite{PhysRevLett.84.1643}. For $\lambda>1$ all the states are
localized in an infinite system and for $\lambda<1$ they are all extended. Fixing the external temperature at $277K$ and the phase breaking at $300 cm^{-1}$ (in spectroscopic wavenumber units) similar to the FMO complex
the parameter $J$ defines the energy scale of the model. For values $J\approx 500-2000 cm^{-1}$ the phase breaking seems to be optimal in a chain of length $N=30$ and 
the best efficiency and the smallest transport time is attained at the critical point of the metal-insulator transition at $\lambda=1$.  Solid lines show transport efficiency for
$J=500,1000$ and $2000$ (blue, red, black respectively). Dashed lines show the transport time for the same cases. The best transport efficiency and shortest transport
time is reached at the critical point between localization and delocalization at $\lambda=1$. The transport time of 20 picoseconds is in accordance with the trapping rate $1/\kappa=1 ps$ and the Boltzmann factor giving probability $\varrho_{rr}=1/20$ at the exit of the chain $n=30$ at $277K$. }
\end{figure}

\end{document}